\documentstyle[prl,aps,twocolumn]{revtex}
\title{Novel Scaling Behavior for the Multiplicity Distribution
under Second-Order Quark-Hadron Phase Transition}
\author{C. B. Yang$^{1,2}$ and X. Cai$^{1, 3}$} 
\address{$^1$ Institute of 
Particle Physics, Hua-Zhong Normal University, 
Wuhan 430079, the People's Republic of China\\
$^2$ Theory Division, RMKI, KFKI,
Budapest 114., Pf. 49, H-1525 Hungary\\
$^3$ Physics Department, Hubei University, Wuhan 430062, the People's
 Republic of China}
\date{\today}
\begin{document}
\maketitle

\vskip 0.5cm
\begin{abstract}
Deviation of the multiplicity distribution $P_q$ in small bin from its
Poisson counterpart $p_q$ is studied within the Ginzburg-Landau description
for second-order quark-hadron phase transition. Dynamical factor $d_q\equiv
P_q/p_q$ for the distribution and ratio $D_q\equiv d_q/d_1$ are defined,
and novel scaling behaviors between $D_q$ are found which
can be used to detect the formation of quark-gluon plasma.  
The study of $d_q$ and $D_q$ is also very interesting for other
multiparticle production processes without phase transition.

{{\bf PACS} number(s): 05.70.Fh, 05.40.+j, 12.38.Mh}
\end{abstract}

\vskip 0.5cm

Multiplicity distribution is one of the most important and most easily
accessible experimental quantities in high energy leptonic and hadronic
collisions. From the well-known KNO scaling and its violation [1, 2] to
the novel scaling form [3] investigated very recently the distribution
shows a lot about the dynamical features for the processes. Local multiplicity
distributions have been studied for many years in terms of a variety of
phase space variables [4], and substantial progress has been made recently
in deriving analytical QCD predictions for those observables [5]. Based
on assuming the validity of the local parton hadron duality hypothesis,
those analytical predictions for
the parton level can be compared to experimental data. A global and local
study of multiplicity fluctuations [6] shows, however, that the theoretical
predictions have significant deviation from experimental data. The significant
deviation of theoretical predictions from experimental data indicates that
we know only a little about multiparticle production processes since the
hadronization process in soft QCD is far from being understood.

In this paper we try to investigate multiplicity distribution in some small
two-dimensional kinetic region (which can be rapidity and transverse momentum
or azimuthal angle, for example) in high energy heavy-ion collision processes.
In such collisions a new matter form, quark-gluon plasma (QGP), may be
formed which subsequently cools and decays into the experimentally observed hadrons,
thus the system undergoes a quark-hadron phase transition. The hadrons produced
in such processes may, in principle, carry some relic information
about their parent state. Thus the investigation of the multiplicity
distribution may be interesting and useful for probing the formation of QGP.
In this paper we are limited to discussing multiplicity distribution under the 
assumption of a second-order phase transition, within the Ginzburg-Landau description
for the phase transition. Within the same description for quark-hadron phase 
transition, the scaled factorial moments are studied by a lot of authors for
second-order [7, 8] and first-order [9-12] phase transitions, and a universal
scaling exponent $\nu\simeq 1.30$ is given in [7,8,11,12].

 It is useful to point out that the study of multiplicity fluctuations in photon
production at the threshold of lasing, which shows a similar type of phase transition
[13], is already in its mature age, although the theory and experiment for a 
quark-hadron phase transition are still in their infancy. As explained in [7-12]
the Ginzburg-Landau model, which has been used in describing superconducting
transition and other macroscopic second-order phase transitions, can also be used
to describe the multiplicity fluctuations in both second- and first-order
phase transitions.
 In [12] multiplicity distributions are studied for both first- and
second-order phase transitions. The authors showed that
for second-order phase transition the probability $P_q$ of finding $q$
hadrons in the small bin under investigation
decreases monotonically with the increase of $q$ regardless of value of bin
width and that for first-order phase transition $P_q$ is a decreasing
function of $q$ for small bin width whereas the shape of the distribution
changes with the increase of bin width. Thus the shape of the
distribution was claimed a tool for telling the order of the phase transition.

In this paper we first show that the criterion in [12]
based on the shape of multiplicity distribution 
for the order of the phase transition
is equivocal. This is easily seen once one considers
the trivial case without dynamical fluctuations. For such a case, the
multiplicity distribution $p_q$ is a Poisson one
\begin{equation}
p_q({\overline s})={{\overline s}^q\over q!}\exp(-{\overline s})\ ,
\end{equation} 
 
\noindent with ${\overline s}$ the mean multiplicity. From this distribution
one has
\begin{equation}
{p_{q+1}\over p_q}={{\overline s}\over q+1}\ .
\end{equation}

\noindent If ${\overline s}<2.0$\ $p_q$ is a monotonically decreasing
function of $q$ whereas $p_q$ changes its shape for ${\overline s}>2.0$.
Using the same parameters as in [12] ${\overline s}$ is calculated and
listed in Table 1. Thus one can see that the shapes of multiplicity
distributions given in [12] are similar to those of Poisson ones.
In real experiments, one can always choose bin width to ensure the
mean multiplicity larger than 2.0, then one cannot tell whether
the distribution is shaped due to statistical fluctuations or
due to the dynamics in the phase transitions. So one cannot give
the order of quark-hadron phase transition just from the general
shape of the distributions, and detailed information is needed. This
result is not surprising, because the non-dominant dynamical fluctuations
can only modify the shape of statistical distribution to some extent
but cannot change its general behavior drastically.

Nevertheless, it should be pointed out that the study of multiplicity
distribution is still very interesting and useful for processes with
the onset of quark-hadron phase transitions. In Ginzburg-Landau theory,
the multiplicity distribution turns out to be a Poisson one if the field
is purely coherent. Conversely, the distribution turns into a negative 
binomial one if the field is completely chaotic. In reality, one can
assume multiplicity production arising from a mixture of chaotic and
coherent fields, so the multiplicity distribution in real processes is not
a Poisson one nor a negative binomial one, and the deviation of the
distribution from a Poisson one is due to dynamical fluctuations. The real
quantity concerned is the deviation of experimental $P_q$ from its
theoretical Poisson counterpart $p_q$. Thus studying
the deviation may reveal features of dynamical mechanism involved. Let
the probability of having $q$ hadrons in a certain bin is $P_q$,
the deviation of $P_q$ from its Poisson counterpart $p_q$ can be measured
by a ratio $d_q=P_q/ p_q$. For the definition of $d_q$ to make sense, it is
necessary to let the mean multiplicity ${\overline s}$ for $P_q$
and $p_q$ be the same. Dynamical fluctuations are shown to be existent
if the ratio is far from 1.0, either much larger or much smaller,
for some $q$. The ratio $d_q$ can be called
dynamical factor, since it is 1.0 unless there are dynamical
fluctuations in the process. In the Ginzburg-Landau description
for second-order phase transition $P_q$ is given by [8]
\begin{equation}
P_q(\delta)=Z^{-1}\int {\cal D}\phi
p_q(\delta^2\mid\phi\mid^2)e^{-F[\phi]}\ ,
\end{equation}

\noindent where $Z=\int{\cal D}\phi e^{-F[\phi]}$ the partition function,
$p_q(\bar s)$ the Poisson distribution with average $\bar s$, and $F[\phi]$
the free energy functional 
\begin{displaymath}
F[\phi]=\delta^2\left[a\mid\phi\mid^2+b\mid\phi\mid^4\right]\ .
\end{displaymath}.

It is instructive to note that a free energy functional with $O(N)$ QCD order
parameter was studied in [14]. The free energy functional used here is different
from that in [14] because of the consideration that we are now only interested in
the final state charged hadrons (most of them are $\pi^\pm$) so that a 
two-component order parameter is enough (which is written as a complex number)
for our purpose. One can see that the functional used here can be derived from that
in [14] by integrating out all other components and neglecting higher order powers
of left components in the exponential. One more simplification used in present
functional is that the derivative term is neglected since former studies (see the
last two papers in [7] for details) find out that the term has little contribution to
the universal exponent which is a measure of the fluctuations. Because of this
simplification, the non-Gaussian functional integral can be treated as a normal
integral and can be evaluated directly.

With the free energy functional above the system is in the plasma state for $a>0$
(the order parameter $\mid\phi_0\mid^2$ corresponding to the minimum of $F[\phi]$
is zero) and in hadron phase for $a<0$ (the order parameter $\mid\phi_0\mid^2>0$).
In real experiments the temperature at which hadrons are emitted from the source
is unknown and may be different from event to event. So we treat $a$ as a free
parameter and discuss only for $a<0$ in the following since in the quark phase
with $a>0$ only a few hadrons can be produced through fluctuations.
From the distribution of Eq. (3) one gets the mean multiplicity for $a<0$
\begin{equation}
{\overline s}=x{J_1(\mid a\mid x)\over J_0(\mid a\mid x)}\ ,
\end{equation}

\noindent with $J_n(\alpha)\equiv\int_0^\infty\,dy\, y^n\exp(-y^2+\alpha y)$,
$x= \delta/\sqrt{b}$ representing the bin width $\delta$, $a\propto
T-T_c$ representing the temperature when the phase transition takes place.
For small phase space bin the mean multiplicity in the bin is proportional
to $x$ thus can be very small. Under such circumstance the distribution must
be concentrated around $P_0$, and both $P_q$ and $p_q$ for $q>1$ must be very small,
so a direct comparison between them could induce large uncertainty. This demands
that the bin width in real experimental analysis should be large enough to ensure
the mean multiplicity in the bin not too small (larger than 0.5, say). Of course,
smaller bins can be used if a precise determination of both $P_q$ and $p_q$ can
be obtained from high statistical experimental data. For zero bin
width the relevant results
are rather sensitive to the cascading production of particles through resonances, so
an extremely small bin width should be avoided.

Because of the normalization of both $P_q$ and $p_q$, the dynamical factor
$d_q$ must be larger than 1.0 for some $q$ and less than 1.0 for some other
$q$ if there exist dynamical fluctuations. One can easily derive
\begin{equation}
d_q(x)={J_q(x(\mid a\mid-1))\over J_0(x\mid a\mid)}{J_0^q(x\mid a\mid)\over
J_1^q(x\mid a\mid)}\exp(\bar s)\ .
\end{equation}

\noindent The dependence of $d_q$ on $q$ for different $x$ and $\mid{a}\mid$
is shown in Fig. 1 by choosing $\mid{a}\mid$=1.0 and 2.0, $-\ln x=-$1.0,
0.0, 1.0, 2.0, 3.0, respectively. From this figure, one can see that
the general shapes of $d_q$ are similar for different choices of
$\mid a\mid$ but depend strongly on the bin width $x$. In detail,
for large $x$ (small $-\ln x$ or high mean multiplicity) $d_1$ is
quite large while $d_{q>1}$ are smaller than 1.0. For small $x$
(large $-\ln x$ or low mean multiplicity), however, $d_1$ is smaller than
but close to 1.0 while $d_{q>1}$ are larger than 1.0, indicating that
two or more particles are more likely to be in the same small bin than
for the pure statistical case. This phenomenon may be associated with
the cluster effect or mini-jets in quark-hadron
phase transition. For small $x$ the values for $d_q$ are independent of
$\mid a\mid$. From Fig. 1 one can also see that the dependence of
$d_q$ on $x$ is quite complicated. For some large $q$ $d_q$ is monotonically
increasing with the decrease of $x$, but for small $q$, $d_q$ first decreases
and then increases with the decrease of $x$. Complicated behaviors 
can be seen for $p_q(x)$, considering the fact that $\bar s$ is an increasing
function of $x$ while $p_q(\bar s)$ changes its behavior at ${\bar s}=q$.
But, the ratio $p_q(x)/p_1(x)$ is an increasing function of $x$ for $q>1$.
Moreover there exists a scaling law between $p_q$
\begin{equation}
{p_q({\overline s})\over p_1({\overline s})}={2^{q-1}\over q!}
\left({p_2({\overline s})\over p_1({\overline s})}\right)^{q-1}\ .
\end{equation}

\noindent Thus it may be more interesting to study the dependence on $x$ of
\begin{equation}
D_q\equiv {d_q\over d_1}={P_q/P_1\over p_q/p_1}
\end{equation}

\noindent instead of $d_q$, and one may expect some scaling behavior
of $D_q$ when the resolution is changed.
Now we turn to study $D_q$ for
second-order quark-hadron phase transition.
If there is no dynamical reason, $P_q=p_q$,
one can see that $D_q$ for all $q$ can
have only one value, 1.0, no matter how large or small the bin width is.
So from the range of values $D_q$ takes, one can evaluate
the strength of dynamical fluctuations.
$D_q$ can be expressed in terms of $J_n(\alpha)$ as
\begin{equation}
\ln D_q=(q-1)\ln {J_0(\mid a\mid x)\over J_1(\mid a\mid x)}+\ln
{J_q(x(\mid a\mid-1))\over J_1(x(\mid a\mid-1))}\ .
\end{equation}

\noindent
Besides $x$, there is in last expression another parameter $\mid a\mid$ 
which is a measure of how far from the critical temperature the hadronization
process occurs and is unknown in current experiments. 
First let us fix $\mid{a}\mid$ to be 1.0. One can immediately
see that, for any $x$, $D_q\propto D_2^{q-1}$, a power law satisfied by
Poisson, binomial and many other distributions in statistics. For any
value of $\mid a\mid$ $D_q$ increases monotonically with the decrease
of $x$. This shows that the dynamical influence can be observed more easily
in high resolution analysis. This can be understood physically, 
since different dynamical fluctuations may offset each other and become less
obviously observable in large bin analysis. The behaviors of $\ln D_q$
as functions of resolution $-\ln x$ are shown in Fig. 2 for $\mid
a\mid$=1.0 and 2.0 for $q$=2,3,4,5,6. For small $-\ln x$ values of
$D_q$ depend strongly on parameter $\mid a\mid$, but they approach
parameter $\mid a\mid$ independent values for large
$-\ln x$. The similarity in the shapes of $\ln D_q$ as functions of $-\ln x$
suggests a power law for other $\mid a\mid$ between $D_q$ and $D_2$ similar to
the case we showed for $\mid a\mid=1.0$. $\ln D_q$ are reshown in Fig. 3 as
functions of $\ln D_2$ with the same data as in Fig. 2. For both
$\mid a\mid$=1.0 and 2.0 perfect linear dependences of $\ln D_q$ on
$\ln D_2$ can be seen. For other values of $\mid a\mid$ the similar linear
dependence is checked to be true. Thus one has
\begin{equation}
\ln D_q=A_q+B_q\ln D_2\ ,
\end{equation}

\noindent with $A_q$ and $B_q$ depending on $\mid a\mid$. The fitted results
of $A_q$ and $B_q$ from curves in Fig. 3 are shown
in Fig. 4 as functions of $\ln (q-1)$ for $\mid a\mid$=1.0 and 2.0. It is
obvious that both $\ln A_q$ and $\ln B_q$ have linear dependence on $\ln (q-1)$ for
fixed $\mid a\mid$. Especially, for the purpose of studying power law,
we investigate $B_q$ and
find that
\begin{equation}
B_q=(q-1)^\gamma\ ,
\end{equation}

\noindent with $\gamma$ depending on $\mid a\mid$. For visualization, the
linear fitting curves for $\ln B_q$ vs $\ln (q-1)$ are shown also
in Fig. 4 for $\mid a\mid$=1.0 and 2.0. The slopes for $\ln A_q$ is about
twice those for $\ln B_q$, and they increase with increasing $\mid a\mid$.
When $\mid a\mid$ is zero, corresponding to the case in which hadrons
are produced exactly at the critical point, numerical results show that
$\gamma$ is 0.819. With the increase of $\mid a\mid$, $\gamma$
increases quickly. For sufficiently large $\mid a\mid$, when the difference
between $\mid a\mid-1$ and $\mid a\mid$ can be neglected, corresponding to
the case in which hadrons are produced at temperature much below the critical
point, one finds that $D_q(x)=F_q(\mid a\mid x)$, with $F_q$ the scaled
factorial moment which is given in [8] for second-order phase transition as
\begin{eqnarray*}
F_q(x)=J_q(x)\,J_0^{q-1}(x)\,J_1^{-q}(x)\ .
\end{eqnarray*}

\noindent Similar relation between $D_q$ and $F_q$ is also true in the small
$x$ limit. In these limiting cases, the scaling of $D_q$ is equivalent to
that of the scaled factorial moments $F_q$, and one can get the exponent
$\gamma$=1.3424 for large $-\ln x$ [15] or large $\mid a\mid$. The dependence
of $\gamma$ on $\mid a\mid$ is shown in Fig. 5. In real experiments,
$\mid a\mid$ is not known for an event and may be increasing in the hadronization
process. Thus some average over $\mid a \mid$ should be made. The smaller
$\mid a\mid$, the less the number of produced particles. Thus the main
contribution to $P_q$ comes from events with large $\mid a\mid$ or with high
multiplicities. For those events, one should get $\gamma$ near 1.30, close
to the universal exponent $\nu$ given in former studies of $F_q$ for
second-order phase transition. For events with low multiplicity, one can
get $\gamma>0.819$. So the theoretical range for the exponent $\gamma$ is
(0.819, 1.3424), corresponding to temperature range from $T=T_C$ to
$T\ll T_C$.

In conclusion, two new quantities $d_q$ and $D_q$ are introduced to
describe dynamical fluctuations in quark-hadron phase transitions.
In the Ginzburg-Landau description for second-order quark-hadron phase
transition, $d_q$ and $D_q$ are investigated analytically, and it is
found that $D_q$ obeys a power law, $D_q\propto
D_2^{B_q}$, with $B_q=(q-1)^\gamma$. In experimental analysis, both $d_q$
and $D_q$ can be obtained quite easily. To get $P_q$ one only needs to count
up the number of events with exact $q$ hadrons in the bin. $p_q$ is of
Poisson type and can be calculated from the experimental $\bar s$. Simple
algebras give $d_q$ and $D_q$. The existence of dynamical fluctuations
can be confirmed if $d_q$ and $D_q$ can take values very
different from 1.0. The scaling between $D_q$ and $D_2$ is a possible
signal for the formation of QGP, because up to now no other dynamical reason 
is known to induce such a scaling. The value of exponent $\gamma$ can be used
to measure the deviation of the temperature, at which the hadronization occurs,
from the critical point. The study of $D_q$ should be carried out in real
experimental analysis in the future to see whether QGP has been formed
in current high energy heavy-ion collisions.     
As a tool to study dynamical fluctuations $d_q$ and $D_q$ introduced in this
paper may also be interesting in experimental analysis of leptonic and hadronic
interactions without quark-hadron phase transitions. Study of $d_q$ and $D_q$
in first-order quark-hadron phase transition is in preparation.

This work was supported in part by the NNSF, the SECF and Hubei NSF in China.

\vskip 1.5cm
\begin{center}\LARGE{\bf TABLE}\end{center}
\begin{table}
\begin{center}
\begin{tabular}{c|c|c|c|c}
$x$ & 1 & 2 & 4 & 6 \\ \hline
$B$=+1 & 0.226 & 0.429 & 0.804 & 1.155\\ \hline
$B=-1$ & 0.342 & 0.797 & 2.172 & 4.582\\ 
\end{tabular}
\caption{Mean multiplicities ${\overline s}$ for second-order ($B$=+1) and
first-order ($B=-1$) phase transitions for different bin widths. $x$ is a
parameter (different from the quantity used in this paper) associated 
with bin width $\delta$, parameter $s$ in [12].
The mean multiplicities are calculated using Eq. (13) in [12].}
\end{center}
\end{table}

\vskip 1cm
\begin{center} \LARGE{\bf Figure Captions}\end{center}
\begin{description}
\item{Fig. 1}\ \ \ Dependence of dynamical factor $d_q$ on $q$ for
$\mid a\mid$=1.0 and 2.0, for $-\ln x=-1.0, 0.0, 1.0, 2.0, 3.0$.
\item{Fig. 2}\ \ \ Dependences of $D_q$ on bin width $-\ln x$ for $\mid a\mid$=
1.0 and 2.0 for $q$=2, 3, 4, 5, 6.

\item{Fig. 3}\ \ \ Scaling behaviors between $D_q$ and $D_2$ for
$\mid a\mid$=1.0 and 2.0. The data are the same as in Fig. 2. From lower
to upper are curves for $q$=3, 4, 5, 6, respectively.

\item{Fig. 4}\ \ \ Coefficients for the scaling between $D_q$ and $D_2$,
$\ln D_q=A_q+B_q\ln D_2$, as 
functions of $\ln(q-1)$ for $\mid a\mid$=1.0 and 2.0.
Linear fitting curves are shown for $B_q=(q-1)^\gamma$.

\item{Fig. 5}\ \ \ Dependence of exponent $\gamma$ on $\mid a\mid$. For large
$\mid a\mid$, $\gamma$ is about 1.34.
\end{description} 
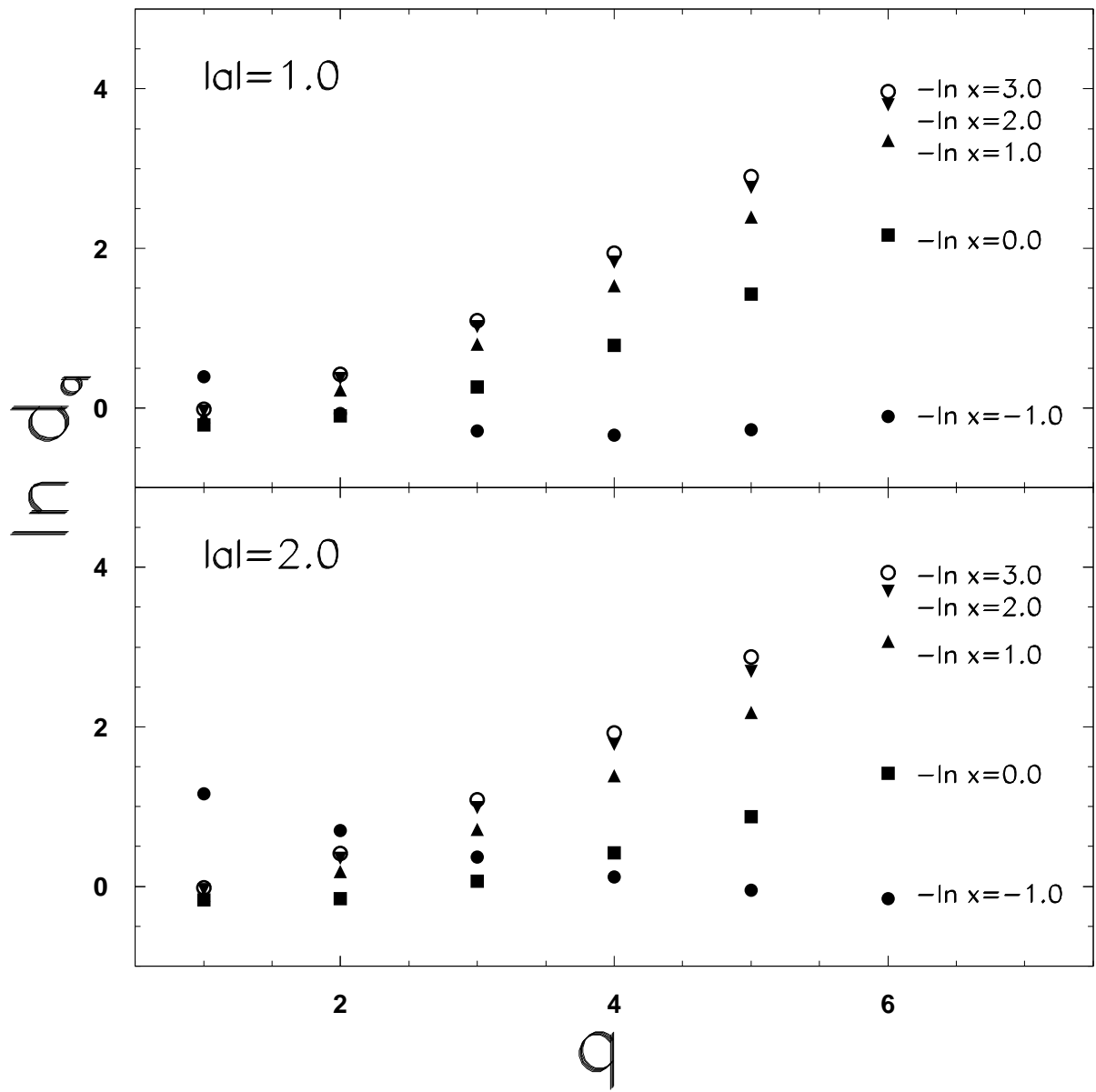
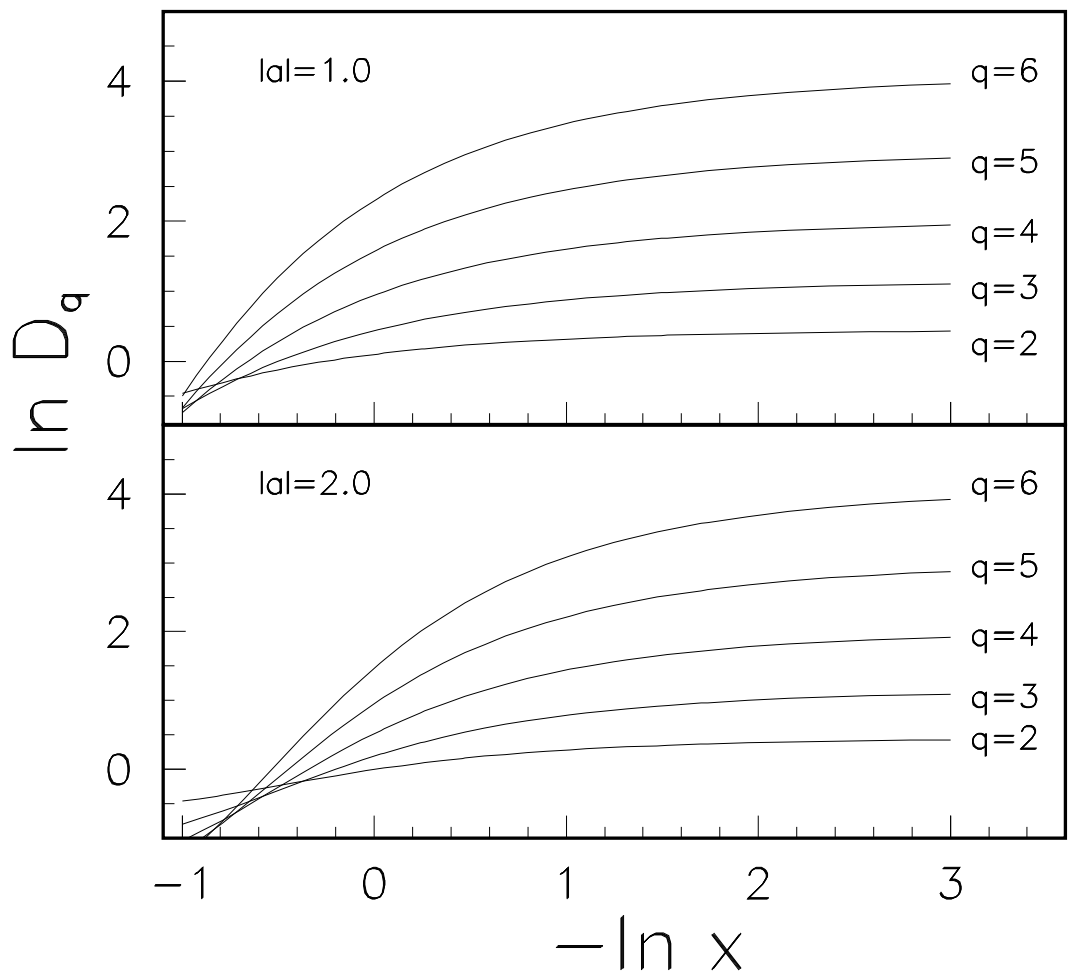
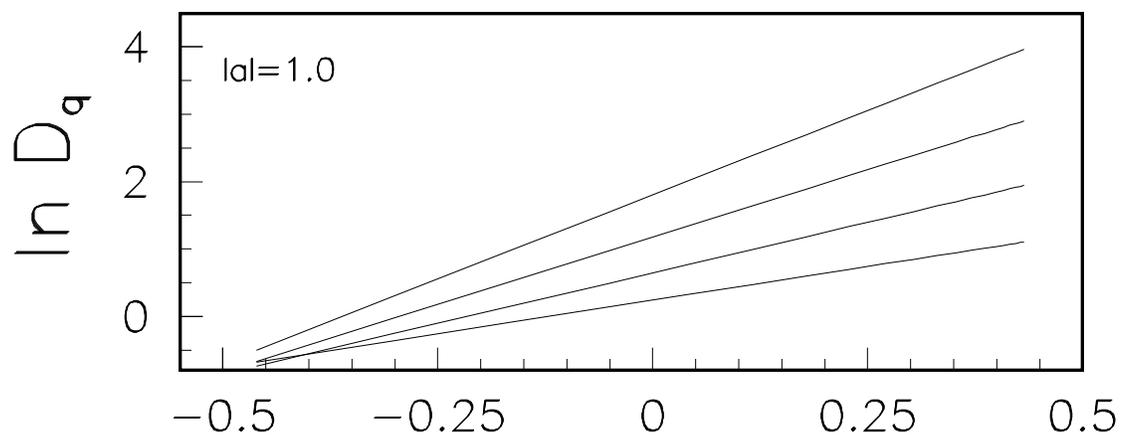
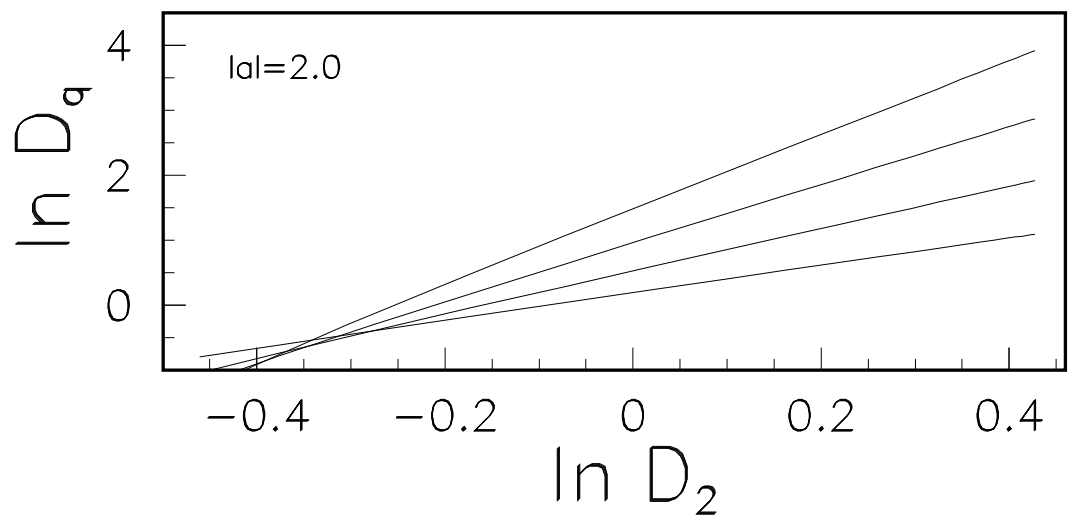
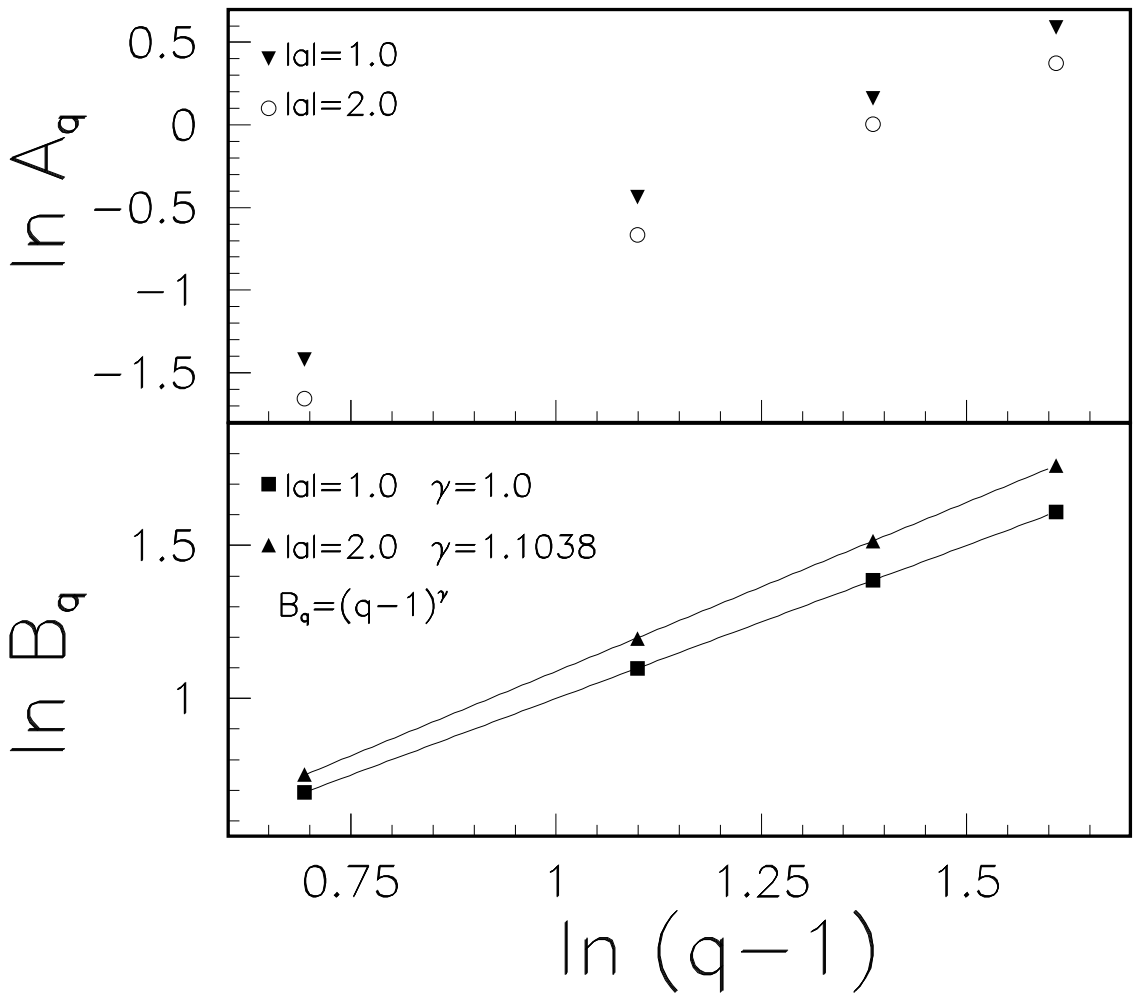
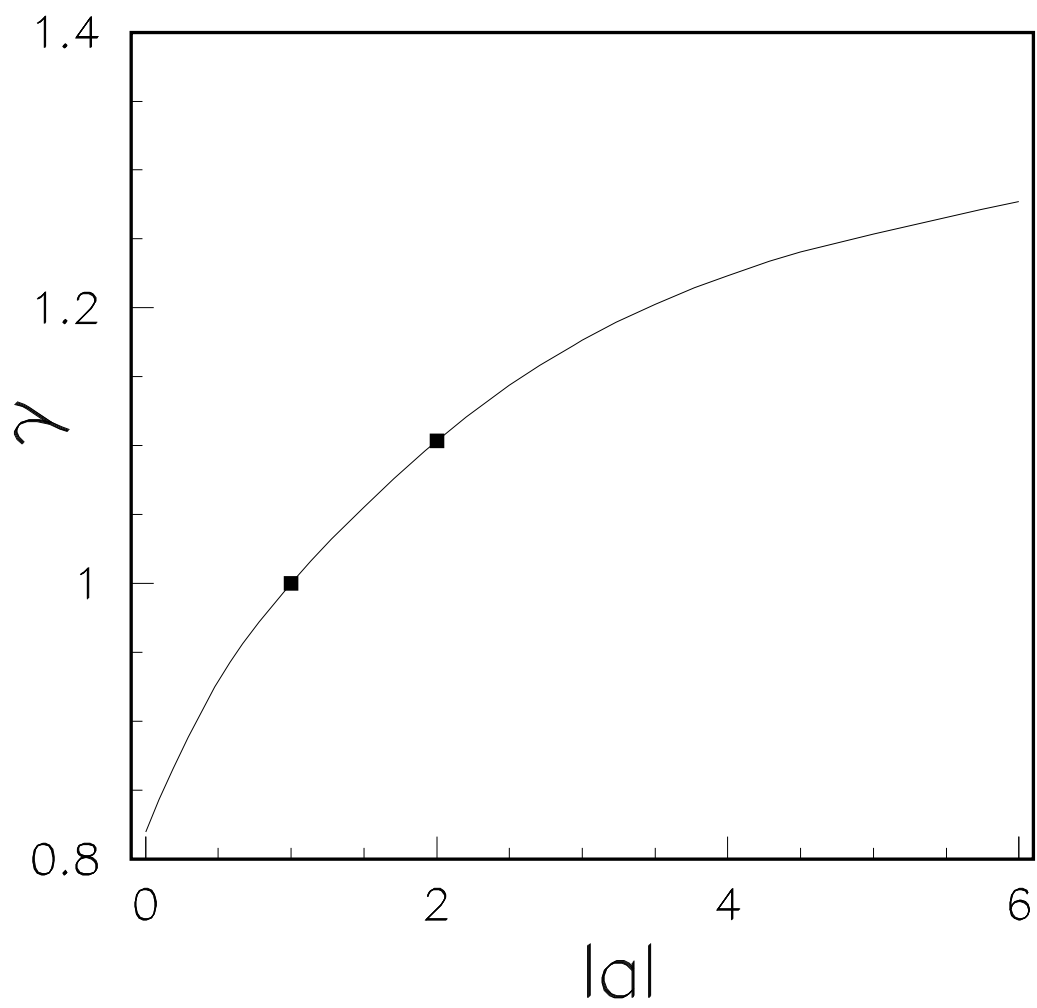
\end{document}